\documentstyle[12pt]{article}

\begin{document}
\rightline{ITP-95-65E}
\bigskip\bigskip
\centerline{\bf HEAT KERNEL EXPANSION FOR OPERATORS OF THE TYPE}
\centerline{\bf OF THE SQUARE ROOT OF THE LAPLACE OPERATOR}
\bigskip\bigskip
\centerline{\sl E.V. Gorbar}
\smallskip
\centerline{\sl Bogolyubov Institute for Theoretical Physics}
\centerline{\sl 252143 Kiev, Ukraine}
\bigskip\bigskip
\centerline{\bf Abstract}
\bigskip
A method is suggested
for the calculation of the DeWitt-Seeley-Gilkey (DWSG) coefficients for the
operator $\sqrt{-\nabla^2 + V(x)}$ basing on a generalization of the
pseudodifferential operator technique.  The lowest DWSG
coefficients for the operator $\sqrt{-\nabla^2} + V(x)$ are calculated by using
the method proposed.  It is shown that the method admits a generalization to
the case of operators of the
type $(-\nabla^2 + V(X))^{1/{\rm m}}$, where m is an arbitrary rational number.
A more simple method is proposed
for the calculation of the DWSG coefficients for the case of strictly
positive operators under the sign of root.  By using this method,
it is shown that the problem of the calculation of the DWSG coefficients for
such operators is exactly solvable.  Namely, an explicit formula expressing
the DWSG coefficients for operators with root through the DWSG coefficients for
operators without root is deduced.
\eject

\section{Introduction}

        The algorithms for obtaining the asymptotic heat kernel expansion
for second order differential operators on a Riemannian manifold are well
known [1-3].  The most popular is that of DeWitt [1,4] which
uses a certain ansatz for heat kernel matrix elements.  The method possesses
the explicit covariance with respect to gauge and general-coordinate
transformations.  However, the DeWitt technique does not apply to
higher-order operators, nonminimal operators and, generally speaking, to
operators whose leading term is not a power of the Laplace operator.  Recently,
using the Widom generalization [5] of the pseudodifferential operator
technique a new algorithm was developed [6] for computing the
DeWitt-Seeley-Gilkey (DWSG) coefficients.  The method is explicitly gauge and
geometrically covariant and admits to carry out calculations of the
DWSG coefficients by computer [7].  As was shown in [8,9], the method permits a
generalization to the case of Riemann-Cartan manifolds, i.e., manifolds with
torsion, and to the case of nonminimal
differential operators.  In this paper the method of ref. [6] is generalized to
the case of operators of the type of the square root of the Laplace operator.
In order that extraction of the root be meaningful, the operator under the
sign of root should be nonnegative, i.e., eigenvalues should be only positive or
zero.  It is not an essential restriction as to the applicability of the method
proposed
because in physics we are mainly encountered with operators bounded from below.
Up to our knoweledge, there is no available
method in current literature for the calculation of the DWSG coefficients for
such type of operators.
In Section 2 we compute the lowest $E_2$ DWSG coefficient for the
operator $\sqrt{-\nabla^2 + V(x)}$ by using a method proposed with the expansion
of the root.  In Section 3 we generalize the method
to the case of an arbitrary natural root, i.e., for the operator
$(\sqrt{-\nabla^2} + V(x))^{1/m}$ where $m$ is any natural and also to case of
the operator
$\sqrt{-\nabla^2} + V(x)$ which cannot be represented as any power of the
operator $-\nabla^2 + V(x)$.  In Section 4, for the case of strictly positive
operators under the sign of root, we propose a more simple method for
the calculation of the DWSG coefficients.
By using this method, we were
able to shown that the problem of the calculation of the DWSG coefficients for
strictly positive operators under the sign of root is exactly solvable.
Namely, an explicit formula expressing the DWSG coeffcients for operators with
root through those for operators without root is deduced.

\section{ Method for calculation of the DWSG coefficients for the
operator of the square root of the Laplace operator}

We take as our space a compact
n-dimensional Riemann manifold $M$ without a boundary.  The operators will act
on the space of a vector bundle over the base $M$.  The covariant derivative
acting on objects with fiber (left understood) and base indices is defined
by the rule
\begin{equation}
\nabla_{\mu}\phi_{\mu_1\dots\mu_k} = (\partial_{\mu} + \omega_{\mu})\phi_{\mu_1
\dots\mu_k} - \sum_{i=1}^{k}\Gamma_{\mu_{i}\mu}^{\lambda} \phi_{\mu_1\dots\mu_
{i-1}\lambda\mu_{i+1}\dots\mu_k}
\end{equation}
where $\Gamma_{\mu \nu}^{\lambda}$ and $\omega_{\mu}$ are the affine and
bundle connections;
$\omega_{\mu} = - \frac{1}{2} i \omega_{\mu}^{ab} \Sigma_{ab} + iA_{\mu}$,
and $\Sigma_{ab}$ are the
representation operators of local rotation group SO(n) under which
$\phi_{\mu_1 \dots \mu_k}$ is
transformed, $\omega_{\mu}^{ab}$ is a spin connection and $A_{\mu}$ is gauge
potential.  For the
commutator of covariant derivatives we have
\begin{equation}
[\nabla_{\mu}, \nabla_{\nu}]\phi_{\mu_1\dots\mu_k} = - \sum_{i=1}^{k}R_{\mu_i\mu
\nu}^{\lambda} \phi_{\mu_i\dots\mu_{i-1}\lambda\mu_{i+1}\dots\mu_k} + W_{\mu\nu}
\phi_{\mu_1\dots\mu_k},
\end{equation}
where $W_{\mu\nu}=\partial_{\mu}\omega_{\nu} -\partial_{\nu}\omega_{\mu} +
[\omega_{\mu},\omega_{\nu}]$ is the bundle
curvature, and the Riemann curvature tensor $R_{\rho \mu \nu}^{\lambda}$ is
expressed through the affine connection $\Gamma_{\mu\nu}^{\lambda}$ as follows:
\begin{equation}
R_{\rho\mu\nu}^{\lambda} = \partial_{\mu}\Gamma_{\rho\nu}^{\lambda} - \partial_
{\nu}\Gamma_{\rho\mu}^{\lambda} + \Gamma_{\sigma\mu}^{\lambda}\Gamma_{\rho\nu}^
{\sigma} - \Gamma_{\sigma\nu}^{\lambda}\Gamma_{\rho\mu}^{\sigma}.
\end{equation}
	As it is well khown [2,3,10], for a positive elliptic differential
operator $A$ of the order $2r$ there exists an asymptotic expansion of the
diagonal matrix elements of the heat kernel ${\rm exp}(-tA)$ as
$t\rightarrow 0_+$ in the following form:
\begin{equation}
<x|e^{-tA}|x>\simeq \sum_{m}E_m(x|A)t^{(m-n)/2r},
\end{equation}
where the summation is carried out over all non-negative integers m and
$E_m(x|A)$ are the DWSG coefficients.

In this section we consider a generalization of the method of [6] to the case of
the operator
\begin{equation}
A = \sqrt{-\nabla^2 + V(x)},
\end{equation}
where V(x) is an arbitrary matrix with respect to bundle space indices.
Following the method [6], to obtain expansion (4) we use the representation of
the operator ${\rm exp}(-tA)\, (t > 0)$ through the operator A resolvent
\begin{equation}
e^{-tA}=\int \limits_{C} \frac{id\lambda}{2\pi} e^{-t\lambda}(A-\lambda)^
{-1},
\end{equation}
where the countour $C$ goes counterclockwise around the spectrum of the
operator $A$.
For the matrix elements of the resolvent $(A - \lambda)^{-1}$ we employ the
representation in the form
\begin{equation}
<x|\frac{1}{A-\lambda}|x^{\prime}>=\int \frac{d^nk}{\left(2\pi\right)^n\sqrt
{g(x^{\prime})}}\, e^{il(x, x^{\prime},k)} \sigma(x, x^{\prime},k;\lambda),
\end{equation}
where $l(x, x^{\prime}, k)$ is a phase function and $\sigma(x, x^{\prime}, k;
\lambda)$ is an amplitude [10, 11].  In the flat space the phase
$l(x, x^{\prime}, k) = k_{\mu}(x - x^{\prime})^{\mu}$ is a linear function of
k and x for each $x^{\prime}$.  In the case of a curved manifold the real
function l must be  biscalar with respect to general-coordinate transformations,
and must be a linear homogeneous function in k.  The generalization of the
linearity condition in x is the requirement for the mth symmetrized covariant
derivative to vanish at the point $x^{\prime}$ with $m \ge 2$, i.e.
\begin{eqnarray*}
\left\{\nabla_{\mu_1}\nabla_{\mu_2}\ldots\nabla_{\mu_m}\right\}l|_{x=x^{\prime}}=
\left[\left\{\nabla_{\mu_1}\nabla_{\mu_2}\ldots\nabla_{\mu_m}\right\}l\right] =
\end{eqnarray*}
\begin{equation}
k_{\mu_1}\,\,\, {\rm for}\ m=1\,\,\, {\rm and}\,\,\, 0\,\,\, {\rm for}\,\,\, m\ne 1.
\end{equation}
In eq. (8) the curly brackets denote symmetrization in all indices and the
square brackets mean that the coincidence limit is taken.  The local properties
of the function l are sufficient to obtain the diagonal heat
kernel expansion.

The resolvent of the operator A satisfies the equation
\begin{equation}
\left(A(x,\nabla_{\mu})-\lambda\right)\Gamma(x,x^{\prime},k;
\lambda)= \frac{1}{\sqrt{g}}\delta (x - x^{\prime}),
\end{equation}
and, therefore, in order to fulfill (9) it is sufficient to require that
the amplitude $\sigma(x,x^{\prime},k;\lambda)$ satisfy the equation
\begin{equation}
\left(A(x,\nabla_{\mu}+i\nabla_{\mu}l)-\lambda\right)\sigma(x,x^{\prime},k;
\lambda)= {\rm I}(x,x^{\prime}).
\end{equation}
The biscalar function ${\rm I}(x, x^{\prime})$ is a matrix with respect to
bundle space indices and is
defined by the conditions similar to eq. (8):
\begin{eqnarray*}
[{\rm I}] = 1,
\end{eqnarray*}
\begin{equation}
\left[\left\{\nabla_{\mu_1}\nabla_{\mu_2}\dots\nabla_{\mu_m}\right\}{\rm I}\right] = 0
\,\,\,m \ge 1,
\end{equation}
the unity in eq. (11) is a matrix unity.
        To generate expansion (4), we introduce an auxiliary parameter
$\epsilon$ into eq. (10) according to the rule
$l \rightarrow l/\epsilon$, $\lambda \rightarrow \lambda / \epsilon,$
and expand the amplitude in a formal series in the powers of $\epsilon$
\begin{equation}
\sigma(x,x^{\prime},k;\lambda)=\sum_{m=0}^{\infty}\epsilon^{1+m}
\sigma_m(x,x^{\prime},k;\lambda)
\end{equation}
(the parameter $\epsilon$ then set equal to one).  Then, eq. (10) gives us
the recursion equations to determine the coefficients $\sigma_m$, and, finally,
this procedure leads to expansion (4) where the DWSG coefficients $E_m(x|A)$ are
expressed through the integrals of $[\sigma_m]$ in the form [6]:
\begin{equation}
E_m(x|A)= \int \frac{d^nk}{(2\pi)^n\sqrt{g}} \int \limits_C \frac{id\lambda}
{2\pi} e^{-\lambda} [\sigma_m](x,x,k;\lambda).
\end{equation}
        Up to now we followed [6] very closely.  Differences arise for the
operator of the type of the square root of the Laplace operator when
we are going to obtain the recursion relations for $\sigma_m$.  For the
ordinary Laplace operator the recursion relations for $\sigma_m$ follow directly
from eq. (10) but it is not a case for the operator
$A = \sqrt{-\nabla^2 + V(x)}$ which we consider.  Explicitly,
the equation for $\sigma$ takes the form
\begin{equation}
(\sqrt{\nabla_{\mu}l\nabla^{\mu}l - i\epsilon\nabla^2l - \epsilon^2\nabla^2 -
2i\epsilon\nabla_{\mu}l\nabla^{\mu} + \epsilon^2V(x)} - \lambda)
\sum_{m=0}^{\infty}\epsilon^m\sigma_m = {\rm I}
\end{equation}
We cannot just expand the square root of the operator in the powers of
$\epsilon$ in the Tailor series as in the case of the Laplace operator because
$\nabla_{\mu}l\nabla^{\mu}l$ and the
operator with $\epsilon$ and $\epsilon^2$ do not commute and it is not
clear in which order to place them
in the Tailor formula.  Therefore, to generate an expansion of the root in
powers of $\epsilon$ we first write down a general structure for
the expansion of the root in the powers of $\epsilon$
\begin{eqnarray*}
\sqrt{\nabla_{\mu}l\nabla^{\mu}l - i\epsilon\nabla^2l - \epsilon^2\nabla^2 -
2i\epsilon\nabla_{\mu}l\nabla^{\mu} + \epsilon^2V(x)} =
\sqrt{\nabla_{\mu}l\nabla^{\mu}l}
+ \epsilon f_1 +
\end{eqnarray*}
\begin{equation}
\epsilon^2 f_2+ \dots + \epsilon^m f_m + \dots
\end{equation}
where $f_{1}$, $f_{2}$ and $f_m$ are to be found.  In order
to show how the method proposed works, first we compute the lowest $E_0$ and
$E_2$ DWSG coefficients.  To do this, we have to find the expansion
of the root up to $\epsilon^2$, i.e. we have to find only $f_1$
and $f_2$.  This can be done as follows:  First,
we take the square of
eq. (15)
\begin{eqnarray*}
{\nabla_{\mu}l\nabla^{\mu}l - i\epsilon\nabla^2l - \epsilon^2\nabla^2 -
2i\epsilon\nabla_{\mu}l\nabla^{\mu} + \epsilon^2V(x)} =
\nabla_{\mu}l\nabla^{\mu}l +
\end{eqnarray*}
\begin{equation}
\sqrt{\nabla_{\mu}l\nabla^{\mu}l} \epsilon f_1 +
\epsilon f_1\sqrt{\nabla_{\mu}l\nabla^{\mu}l} + \epsilon^2 f_1^2 +
\sqrt{\nabla_{\mu}l\nabla^{\mu}l} \epsilon^2 f_2 +
\epsilon^2 f_2 \sqrt{\nabla_{\mu}l\nabla^{\mu}l} + \dots
\end{equation}
Then, comparing terms with the equal powers of $\epsilon$, we obtain
the equations for $f_1$ and $f_2$
\begin{equation}
-i\nabla^2l - 2i\nabla_{\mu}l \nabla^{\mu} =
\sqrt{\nabla_{\mu}l\nabla^{\mu}l} f_1 +
f_1\sqrt{\nabla_{\mu}l\nabla^{\mu}l},
\end{equation}
\begin{equation}
-\nabla^2 + V(x) = f_1^2 +
\sqrt{\nabla_{\mu}l\nabla^{\mu}l} f_2 +
f_2 \sqrt{\nabla_{\mu}l\nabla^{\mu}l}
\end{equation}
From the left-hand side of eq. (17) it follows that a general structure
of $f_1$ is $f_1 = -ia_{\mu}\nabla^{\mu} - ib$, where $a_{\mu}$ and b are
ordinary vector and scalar functions, respectively, not operators.
Substituting the general expression for $f_1$ into eq. (17), we get
the following equations for $a_{\mu}$ and b:
\begin{equation}
-2i\nabla_{\mu}l\nabla^{\mu} = -i
\sqrt{\nabla_{\nu}l\nabla^{\nu}l} a_{\mu}\nabla^{\mu} -
ia_{\mu}\sqrt{\nabla_{\nu}l\nabla^{\nu}l}\nabla^{\mu}
\end{equation}
\begin{equation}
-i\nabla^2l = -i\sqrt{\nabla_{\mu}l\nabla^{\mu}l}b -
ib\sqrt{\nabla_{\mu}l\nabla^{\mu}l} -
ia_{\mu}\nabla^{\mu}\sqrt{\nabla_{\nu}l\nabla^{\nu}l}
\end{equation}
From these equations we obtain
\begin{eqnarray*}
a_{\mu} = \frac{\nabla_{\mu}l}
{\sqrt{\nabla_{\nu}l\nabla^{\nu}l}},
\end{eqnarray*}
\begin{equation}
b= \frac{\nabla^2l}{2\sqrt{\nabla_{\mu}l\nabla^{\mu}l}} -
\frac{\nabla_{\mu}l\nabla^{\mu}\sqrt{\nabla_{\nu}l\nabla^{\nu}l}}
{2\nabla_{\alpha}l\nabla^{\alpha}l}
\end{equation}
Finally, the equation for $f_{2}$ takes the form
\begin{equation}
-\nabla^2 + V(x) = - (a_{\mu}\nabla^{\mu} + b)^2 +
\sqrt{\nabla_{\mu}l\nabla^{\mu}l} f_2 +
f_2 \sqrt{\nabla_{\mu}l\nabla^{\mu}l}
\end{equation}
Similarly to the case of $f_1$ we write down a general structure of $f_2$
\begin{equation}
f_2 = C_{1}\nabla^2 + C_{2\mu}\nabla^{\mu} +
C_{3\mu \nu}\nabla^{\mu}\nabla^{\nu} + C_4
\end{equation}
Substituting (22) into eq.(24), we find
\begin{eqnarray*}
C_1 = -\frac{1}{2R^{1/2}},
\end{eqnarray*}
\begin{eqnarray*}
C_{2\mu} =
\frac{\nabla_{\mu}R^{1/2}}{2R} -
\frac{a_{\mu}a_{\nu}\nabla^{\nu}R^{1/2}}{2R} +
\frac{a_{\mu}b}{R^{1/2}} +
\frac{a_{\nu}\nabla^{\nu}a_{\mu}}{2R^{1/2}},
\end{eqnarray*}
\begin{eqnarray*}
C_{3\mu \nu} = \frac{a_{\mu}{a_\nu}}{2R^{1/2}},
\end{eqnarray*}
\begin{eqnarray*}
C_4 = \frac{V(x)}{2R^{1/2}} +
\frac{a_{\mu}\nabla^{\mu}b}{R^{1/2}} + \frac{b^2}{2R^{1/2}} -
\frac{\nabla^2R^{1/2}}{4R} -
\frac{1}{2R^{1/2}}(\frac{\nabla_{\mu}R^{1/2}\nabla^{\mu}R^{1/2}}{2R} -
\end{eqnarray*}
\begin{equation}
\frac{a_{\nu}a_{\mu}\nabla^{\nu}R^{1/2}\nabla^{\mu}R^{1/2}}{2R} +
\frac{a_{\mu}b\nabla^{\mu}R^{1/2}}{R^{1/2}} +
\frac{a_{\nu}\nabla^{\nu}a_{\mu}\nabla^{\mu}R^{1/2}}{2R^{1/2}}) -
\frac{a_{\mu}a_{\nu}\nabla^{\mu}\nabla^{\nu}R^{1/2}}{2R},
\end{equation}
where $R=\nabla_{\mu}l\nabla^{\mu}l$.

Thus, we have found the expansion of the root up to $\epsilon^2$ but it
is obviously that it is possible in a similar way to find the expansion of the
root up to any mth
power of $\epsilon$ because the equation for $f_m$ has a similar form
to the equations for $f_1$ and $f_2$, namely,
$\sqrt{\nabla_{\mu}l\nabla^{\mu}l} f_m + f_m \sqrt{\nabla_{\mu}l\nabla^{\mu}l}$.
Consequently,
writting down a general stucture of $f_m$ and defining $f_1$, $f_2$,...,
$f_{m-1}$, we can find the explicit expression for $f_m$ in the same way as
it was done in the case of $f_1$ and $f_2$.

Having obtained the explicit expansion of the root
\begin{eqnarray*}
\sqrt{\nabla_{\mu}l\nabla^{\mu}l - i\epsilon\nabla^{2}l - \epsilon^2\nabla^2 -
2i\epsilon\nabla_{\mu}l\nabla^{\mu} + \epsilon^{2}V(x)} =
\end{eqnarray*}
\begin{equation}
\sqrt{\nabla_{\mu}l\nabla^{\mu}l} -
i\epsilon(a_{\mu}\nabla^{\mu} + b)  +\epsilon^2 f_2 + \dots,
\end{equation}
from eq.(14) we have the following equations for $\sigma_0$, $\sigma_1$
and $\sigma_2$:
\begin{eqnarray*}
(R^{1/2} - \lambda)\sigma_0 = {\rm I},
\end{eqnarray*}
\begin{eqnarray*}
(R^{1/2} - \lambda)\sigma_1 -i(a_{\mu}\nabla^{\mu} + b)\sigma_0 = 0,
\end{eqnarray*}
\begin{equation}
(R^{1/2} - \lambda)\sigma_2 -i(a_{\mu}\nabla^{\mu} + b)\sigma_1 +
(C_1\nabla^2 + C_{2\mu}\nabla^{\mu} +
C_{3\mu\nu}\nabla^{\mu}\nabla^{\nu} +
C_4)\sigma_0 = 0.
\end{equation}
From eqn.(26) we obtain
\begin{equation}
[\sigma_0] = \frac{1}{\sqrt{k^2} - \lambda},
\end{equation}
\begin{eqnarray*}
[\sigma_2] = -\frac{k_{\mu}k^{\lambda}l^{\mu\nu}_{\,\,\,\,\nu\lambda}}{2k^2(\sqrt{k^2}
-\lambda)^3} -
\frac{k_{\mu}k_{\lambda}l^{\nu\,\,\mu\lambda}_{\,\,\nu}}{2k^2(\sqrt{k^2} - \lambda)^3}
-\frac{V(x)}{2\sqrt{k^2}(\sqrt{k^2} - \lambda)^2} -
\end{eqnarray*}
\begin{equation}
\frac{k_{\mu}k^{\lambda}l^{\mu\nu}_{\,\,\,\,nu\lambda}}{4(k^2)^{3/2}
(\sqrt{k^2} - \lambda)^3} -
\frac{k_{\mu}k_{\lambda}l_{\,\,\nu}^{\nu\,\,\,\,\mu\lambda}}{4(k^2)^{3/2}
(\sqrt{k^2} - \lambda)^3},
\end{equation}
where we introduced the notation $[\nabla_{\mu}\nabla_{\nu}\dots\nabla_{\lambda}
l]=k_{\alpha}l_{\mu\nu\ldots\lambda\alpha}$ (see [6]) and
wrote down only terms which do not vanish after the substitution
of the explicit expression for $l_{\mu\nu\ldots\lambda}$ and the convolution with
$k^{\mu}k^{\nu}\ldots k^{\lambda}$.

Let us recall that the DWSG coefficients are given by (13) and we have to
calculate the integrals in $\lambda$ and k.  The integral in $\lambda$ is
trivially calculated by using the residue theory,
and, consequently, we obtain
\begin{equation}
E_0(x) = \int\frac{d^nk}{(2\pi)^n\sqrt{g(x)}} {\rm e}^{-\sqrt{k^2}},
\end{equation}
\begin{eqnarray*}
E_2(x) = \int\frac{d^nk}{(2\pi)^n\sqrt{g(x)}} {\rm e}^{-\sqrt{k^2}}
(-\frac{k_{\mu}k^{\lambda}l^{\mu\nu}_{\,\,\,\,\nu\lambda}}{4k^2} -
\end{eqnarray*}
\begin{equation}
\frac{k_{\mu}k_{\lambda}l^{\,\,\nu\mu\lambda}_{\nu}}{4k^2} -
\frac{V(x)}{2\sqrt{k^2}} -
\frac{k_{\mu}k^{\lambda}l^{\mu\nu}_{\,\,\,\,\nu\lambda}}{4(k^2)^{3/2}} -
\frac{k_{\mu}k_{\lambda}l_{\nu}^{\,\,\,\,\nu\mu\lambda}}{4(k^2)^{3/2}}).
\end{equation}
To calculate the integral in k, we note that (see [6])
\begin{eqnarray*}
\int\frac{d^nk}{(2\pi)^n\sqrt{g}}k_{\mu_1}k_{\mu_2}\ldots k_{\mu_{2s}}f(k^2) =
\end{eqnarray*}
\begin{eqnarray*}
g_{\left\{\mu_1\mu_2\ldots\mu_{2s}\right\}} \frac{1}{(4\pi)^{n/2}2^s\Gamma(n/2
+ s)} \int_{0}^{\infty}dk^2(k^2)^{(n-2)/2 + s} f(k^2),
\end{eqnarray*}
\begin{equation}
k^2 = g^{\mu\nu}k_{\mu}k_{\nu},
\end{equation}
where $g_{\left\{\mu_1\mu_2\ldots\mu_{2s}\right\}}$ is the symmetrized sum of
metric tensor products.
Integrating in $k^2$, we obtain
\begin{equation}
E_0(x) = \frac{2\Gamma(n)}{(4\pi)^{n/2}\Gamma(n/2)},
\end{equation}
\begin{eqnarray*}
E_2(x) = \frac{1}{(4\pi)^{n/2}}
(\frac{l^{\mu\nu}_{\nu\mu}\Gamma(n)}{4\Gamma(n/2+1)} -
\frac{V(x)\Gamma(n-1)}{\Gamma(n/2)} -
\frac{l^{\mu\nu}_{\mu\nu}\Gamma(n)}{4\Gamma(n/2+1)} -
\end{eqnarray*}
\begin{equation}
\frac{l^{\mu\nu}_{\nu\mu}\Gamma(n-1)}{4\Gamma(n/2+1)} -
\frac{l^{\mu\nu}_{\mu\nu}\Gamma(n-1)}{4\Gamma(n/2+1)}).
\end{equation}
Using $l_{\mu\nu\lambda\alpha} = -
\frac{R_{\alpha\lambda\mu\nu}}{3} -
\frac{R_{\alpha\nu\mu\lambda}}{3}$ [6], we obtain
\begin{equation}
E_2(x) = \frac{2\Gamma(n-2)}{(4\pi)^{n/2}\Gamma(n/2-1)}\left(\frac{R}{6} - V(x)
\right).
\end{equation}
Note that $E_2$ obtained for the operator of the type of the square root of
the Laplace operator coincides with $E_2$ calculated for the Laplace
operator up to a contant factor and, thus, the dependence of this coefficient on
the space dimension is rather trivial
(cf. with the case of nonmiminal operators [8] whose leading coefficients also
are not a power of the Laplace operator).  We will show in Section 4 that
for strictly positive operators the same is true for the DWSG coefficients of
an arbitrary order.  It is an interesting problem whether it is also true for
operators which have zero eigenmodes.
The method proposed permits a generalization to the case of any rational
root and can be also used for the calculation of the DWSG coefficients for
the operator $\sqrt{-\nabla^2} + V(x)$ whose the square is not the Laplace
operator.

\section{ A generalization to the case of an arbitrary
rational root and the operator which cannot be presented as a power
of the Laplace operator}

In this section we generalize the method proposed to the case
of an arbitrary rational root, i.e. for the operator
$(-\nabla^2 + V(x))^{p/m}$, where p and m are any naturals.  For the sake of
simplicity,
we actually consider the case of a natural root, i.e., the operator of the type
$(-\nabla^2 + V(x))^{1/m}$, where m is any natural number and show that the
method can be easy generalized to the case of an arbitrary rational root.
The equation for $\sigma$ has the form
\begin{equation}
\left((-(\nabla_{\mu}+i\nabla_{\mu}l)(\nabla^{\mu} + i\nabla^{\mu}l)+
V(x))^{1/m}-
\lambda\right)\sigma(x,x^{\prime},k;\lambda)= {\rm I}(x,x^{\prime}).
\end{equation}
To generate the heat kernel expansion, we introduce an auxiliary
parameter $\epsilon$ into eq. (35) according to
the rule $l \rightarrow l/\epsilon^{m/2}$, $\lambda \rightarrow \lambda /
\epsilon$
and expand the amplitude in a formal series in the powers of $\epsilon$
\begin{equation}
\sigma(x,x^{\prime},k;\lambda)=\sum_{s=0}^{\infty}\epsilon^{1+
s\frac{m}{2}}\sigma_s(x,x^{\prime},k;\lambda).
\end{equation}
We seek an expansion of the root in the form
\begin{eqnarray*}
(\nabla_{\mu}l\nabla^{\mu}l - i\epsilon^{m/2}\nabla^2l -
\epsilon^m\nabla^2 - 2i\epsilon^{m/2}\nabla_{\mu}l\nabla^{\mu} +
\end{eqnarray*}
\begin{equation}
\epsilon^mV(x))^{1/m} =
(\nabla_{\mu}l\nabla^{\mu}l)^{1/m}
+ \epsilon^{m/2} f_1 +\epsilon^m f_2+ \dots
\end{equation}
Taking the mth power of eq. (37), we obtain the following equations for
the unknown $f_1$ and $f_2$:
\begin{equation}
-i\nabla^2l - 2i\nabla_{\mu}l \nabla^{\mu} =
\sum_{i=0}^{m-1} R^i f_1 R^{m-1-i},
\end{equation}
\begin{equation}
-\nabla^2 + V(x) = \sum_{i=0}^{m-2}\sum_{j=i}^{m-2} R^i f_1 R^{j/2} f_1 R^{m-
2-i-j} +
\sum_{i=0}^{m-1} R^i f_2 R^{m-1-i},
\end{equation}
where $R=(\nabla_{\mu}l\nabla^{\mu}l)^{1/m}$.  Note that in the case of an
arbitrary ratioanl root, i.e., for the operators of the type
$(-\nabla^2 + V(x))^{p/m}$, we would be have the pth power of
$-i\nabla^2l - 2i\nabla_{\mu}l \nabla^{\mu}$
on the left-hand side of equation (37) that does not present an untractable
problem for generating the expansion in the powers of $\epsilon$.  All the same
we again can write down a general structure in derivatives of the root and
taking the mth power can find the unknowns in the expansion of the root.
Thus, the method can be used in the case of an arbitrary rational root.
Writting down general structures of $f_1$ and $f_2$
\begin{eqnarray*}
f_1 = -i(a_{\mu}\nabla^{\mu} + b),
\end{eqnarray*}
\begin{equation}
f_2 = C_1\nabla^2 + C_{2\mu}\nabla^{\mu} + C_{3\mu\nu}\nabla^{mu\nu} +
C_4,
\end{equation}
from eqn. (38) and (39) we find the explicit expressions for  $f_1$ and $f_2$
and from eq. (35) obtain the recursion relations for $\sigma_0$, $\sigma_1$
and $\sigma_2$
\begin{eqnarray*}
(R - \lambda)\sigma_0 = I,
\end{eqnarray*}
\begin{eqnarray*}
(R - \lambda)\sigma_1 -i(a_{\mu}\nabla^{\mu} + b)\sigma_0 = 0,
\end{eqnarray*}
\begin{eqnarray*}
(R- \lambda)\sigma_2 -i(a_{\mu}\nabla^{\mu} + b)\sigma_1 +
(C_1\nabla^2 + C_{2\mu}\nabla^{\mu} +
\end{eqnarray*}
\begin{equation}
C_{3\mu\nu}\nabla^{\mu}\nabla^{\nu} +
C_4)\sigma_0 = 0,
\end{equation}
where $R = (\nabla_{\mu}l\nabla^{\mu}l)^{1/m}$.
Taking the coincidence limits, we have
\begin{eqnarray*}
[\sigma_0] = \frac{1}{(k^2)^{1/m} - \lambda},
\end{eqnarray*}
\begin{eqnarray*}
[\sigma_2] =
-\frac{2k_{\mu}k^{\lambda}l^{\mu\nu}_{\,\,\,\,\nu\lambda}}{m^2
(k^2)^{2-2/m}((k^2)^{1/m} -\lambda)^3} -
-\frac{ 2 k_{\mu} k_{\lambda} l_{nu}^{\,\,\nu \mu \lambda}}
      {m^2 (k^2)^{2-2/m}( (k^2)^{1/m} - \lambda)^3} -
\end{eqnarray*}
\begin{eqnarray*}
\frac{V(x)}{m(k^2)^{1-1/m}((k^2)^{1/m} - \lambda)^2} -
\end{eqnarray*}
\begin{equation}
\frac{(m-1)k_{\mu}k^{\lambda} l^{\mu \nu}_{\,\,\,\,\nu \lambda}}
     {m^2 (k^2)^{2-1/m} ((k^2)^{1/m} - \lambda)^3} -
\frac{ (m-1) k^{\mu} k^{\lambda} l_{\nu}^{\,\,\nu \mu \lambda}}
     {m^2(k^2)^{2-1/m}((k^2)^{1/m} - \lambda)^3}
\end{equation}
where we write down only terms which do not vanish after the substitution
of the explicit expression for $l_{\mu\ldots\lambda}$ and the convolution with
$k^{\mu\ldots\lambda}$.

Integrating in $\lambda$ and k, we obtain the lowest DWSG coefficients
of the heat kernel expansion for the operator $(-\nabla^2 + V(x))^{1/m}$
\begin{equation}
E_0(x) = \frac{m\Gamma(\frac{m}{2}n)}{(4\pi)^{n/2}\Gamma(n/2)},
\end{equation}
\begin{equation}
E_2(x) = \frac{m\Gamma\left(m(\frac{n-m}{2}\right)}
{(4\pi)^{n/2}\Gamma(\frac{n-m}{2})}\left(\frac{R}{6} - V(x)\right).
\end{equation}
$E_0$ and $E_2$ calculated in the case of an arbitrary natural
root m coincide with $E_0$, $E_2$ calculated in the particular case m=2
(see eqn. (32) and (34)).  Note that in comparision with the case of
nonmiminimal operators [9] the dependence on m and the dimension of space is
rather trivial, namely, $E_2$ depends on m only through a constant factor.

We now show how the method proposed works in the case of the calculation of
the DWSG coefficients for the operator $\sqrt{-\nabla^2} + V(x)$ which cannot be
obviously represented as a power of the Laplace operator.
We can use the old expansion (25) for the root
$\sqrt{\nabla_{\mu}l\nabla^{\mu}l - i\epsilon\nabla^2l - \epsilon^2\nabla^2 -
2i\epsilon\nabla_{\mu}l\nabla^{\mu}}$.  Further,
as usually, in order to generate the heat kernel expansion,
we introduce an auxiliary parameter $\epsilon$ according to
the rule $l \rightarrow l/\epsilon$, $\lambda \rightarrow \lambda / \epsilon$
and expand the amplitude $\sigma$ in a formal series in powers of $\epsilon$
\begin{equation}
\sigma_{\epsilon}(x,x^{\prime},k;\lambda)=\sum_{m=0}^{\infty}\epsilon^{1+m}
\sigma_m(x,x^{\prime},k;\lambda).
\end{equation}
Then, the equations for $\sigma_0, \sigma_1, \sigma_2$ take the form
\begin{eqnarray*}
(R^{1/2} - \lambda)\sigma_0 = I,
\end{eqnarray*}
\begin{eqnarray*}
(R^{1/2} - \lambda)\sigma_1 + (-i(a_{\mu}\nabla^{\mu} + b) +V(x))\sigma_0 = 0,
\end{eqnarray*}
\begin{eqnarray}
&&(R^{1/2} - \lambda)\sigma_2 + (-i(a_{\mu}\nabla^{\mu} + b) +V(x))\sigma_1
\nonumber\\
&&+(C_1\nabla^2 + C_{2\mu}\nabla^{\mu} + C_{3\mu\nu}\nabla^{\mu}\nabla^{\nu} +
C_4)\sigma_0 = 0.
\end{eqnarray}
From eqn.(46) we find
\begin{eqnarray*}
[\sigma_1] = - \frac{V(x)}{(\sqrt{k^2} - \lambda)^2},
\end{eqnarray*}
\begin{equation}
[\sigma_2]_{new} = \frac{k_{\mu}\nabla^{\mu}V(x)}{(\sqrt{k^2} - \lambda)^3
\sqrt{k^2}} + \frac{V^2(x)}{(\sqrt{k^2} - \lambda)^3},
\end{equation}
where we write down only the terms with $V(x)$; the terms with
$l_{\mu\dots\alpha}$ coincide with that for the operator
$\sqrt{-\nabla^2 + V(x)}$.
The first term in the expression for $[\sigma_2]_{new}$ vanishes after the
integration in k because of an odd power of k.  Note also that in difference to
the case of the operator $\sqrt{-\nabla^2 + V(x)}$
coefficient $E_1$ is not equal to zero for the operator $\sqrt{-\nabla^2} +
V(x)$.

Integrating in $\lambda$ and k, we obtain
\begin{equation}
E_1(x) = - \frac{2\Gamma(n)V(x)}{(4\pi)^{n/2}\Gamma(n/2)},
\end{equation}
\begin{equation}
E_2(x)_{new} = - \frac{2\Gamma(n)V^2(x)}{(4\pi)^{n/2}\Gamma(n/2)}.
\end{equation}
Consequently, the entire $E_2$ coefficient is
\begin{equation}
E_2(x)= \frac{\Gamma(n-1)}{(4\pi)^{n/2}\Gamma(n/2)}(\frac{R}{6} -
2(n-1)V^2(x)).
\end{equation}
Note that contrary to the case of the operator $\sqrt{-\nabla^2 + V(x)}$ the
$E_2$ coefficient for the operator $\sqrt{-\nabla^2} + V(x)$ essentially
depends on the dimension of space.
We can also generalize the method proposed to the case of the operator of
the type $(-\nabla^2)^{1/m} + V(x)$, where m is any natural number.
Thus, we have shown that the method proposed can be modified and adopted for
the calculation of the DWSG coefficients for various
operators which involve the extraction of root and have calculated lowest $E_2$
coefficient for three different operators. Amount of work needed to
calculate the DWSG coefficients icreases very quickly with the growth
of the order of the DWSG coeffcient in the method proposed.  It is connected
with rapid increase of the number of terms in the expansion of the root
with the growth of the order of the DWSG coefficient.  In fact, in the case
where the operator under the sign of the root is strictly positive there exists
a more simple and less laborious method for the calculation of the DWSG
coefficients.

\section{More simple method for the calculation of the DWSG coefficients
for strictly positive operators}

Let us again consider the operator $A =\sqrt{-\nabla^2 + V(x)}$.  If this
operator is strictly positive, i.e., it does not have any zero
eigenmodes, we can used instead of (6) the following representation:
\begin{equation}
e^{-tA}=\int \limits_{C} \frac{id\lambda}{2\pi} e^{-t\lambda^{1/2}}(A-\lambda)^
{-1},
\end{equation}
We demand that $-\nabla^2 + V(x)$ do not have any zero eigenmodes because in
such a case the contour $C$ can be drawn such that it encircles the whole
spectrum of the operator $-\nabla^2 + V(x)$ and do not intersect anywhere
the cut from infinity to zero along the negative half-axis which is needed in
order that the extraction of
root be meaningful.   If the operator has zero eigenmodes, such a countour
cannot be drawn because in this case we cannot draw the countour in such a way
that it do not
intersect the cut and simultaneously the contribution of eigenmodes be properly
taken into account.  By using this method, we can prove that the DWSG
coefficients for operators with root are expressed through the DWSG coefficients
for operators without root.  Note that this method cannot be used
for operators which have zero eigenmodes.  However, the method with the
expansion of root can be use also in the case of operators with zeromodes.
Of course, for strictly positive operators both methods
can be used and they yield coinciding results.

$E_m$ coefficients in the method with the
representation $e^{-t\lambda^{1/2}}$ are given by the relation
\begin{equation}
E_m(x|A)= \int \frac{d^nk}{(2\pi)^n\sqrt{g}} \int \limits_C \frac{id\lambda}
{2\pi} e^{-\lambda^{1/2}} [\sigma_m](x,x,k;\lambda),
\end{equation}
where $\sigma_m(x,x,k;\lambda)$ are the same as for the operator
$-\nabla^2 + V(x)$.
Using $\sigma_2$ obtained in work [6] and calculating the integrals over
$\lambda$ and k,
we obtain the following $E_2$ coefficient for
the operator $\sqrt{-\nabla^2 + V(x)}$:
\begin{equation}
E_2(x) = \frac{2\Gamma(n-2)}{(4\pi)^{n/2-1}\Gamma(n/2)}\left(\frac{R}{6} - V(x)
\right)
\end{equation}
which coincides with $E_2$ obtained in Section 2 by using the method with
the expansion of root.

Let us show by using the method with representation (51) that the DWSG
coefficients for operators with root are explicitly expressed through the DWSG
coefficients for operators without root.  Let us consider the DWSG coefficients
for the operator $-\nabla^2 + V(x)$.  According to [6], they
are given by the relation
\begin{equation}
E_m(x|A)= \int \frac{d^nk}{(2\pi)^n\sqrt{g}} \int \limits_C \frac{id\lambda}
{2\pi}
e^{-\lambda} [\sigma_m](x,x,k;\lambda).
\end{equation}
Comparing it with (52),
we see that the only difference is the power of $\lambda$ in the
exponent, namely, it is equal to 1/2 in the case of the operator with root and 1
for the operator without root.  We recall that a general
term of $[\sigma(x,x,k;\lambda)]$ has the following form:
\begin{equation}
\frac{k_{\mu_1}\ldots k_{\mu_{2s}}F^{\mu_1\ldots\mu_{2s}}}{(k^2 -
\lambda)^a},
\end{equation}
where $F^{\mu_1\ldots\mu_{2s}}$ is expressed through the bundle
curvature
$W_{\mu\nu}$ and the Riemannian curvature tensor $R^{\lambda}_{\rho\mu\nu}$.
$[\sigma_m(x,x,k;\lambda)]$ is the sum of terms with various powers of a and
s.
It is very important for what follows that the difference $a - s$ is fixed
for the DWSG coefficient of a given order.  This fact follows from the
homogeneity property of the reccurent relations for $\sigma_m$ (see [6]).
$a - s$ is equal to $1 + m/2$ for the operator $-\nabla^2 + V(x)$.
Integrating over $\lambda$ and angles in n-dimensional space, we have for
the DWSG coefficient in the case of the operator without root
\begin{equation}
\int dk k^{n - 1 + 2s} g_{\{\mu_1\ldots\mu_{2s}\}}F^{\mu_1\ldots\mu_{2s}}
\frac{d^{a - 1}}{d k^{2(a - 1)}}e^{-k^2}
\end{equation}
and for the operator with root
\begin{equation}
\int dk k^{n - 1 + 2s} g_{\{\mu_1\ldots\mu_{2s}\}}F^{\mu_1\ldots\mu_{2s}}
\frac{d^{a - 1}}{d k^{2(a - 1)}}e^{-k},
\end{equation}
where we have omitted common constant factors which coincide for two cases under
consideration and have used formula (31).  Integrating over k, we obtain
$\Gamma(\frac{n-2}{2} + s - a + 2) = \Gamma(\frac{n-m}{2})$ for the operator
without root and $2\Gamma(n-2 + 2s - 2a + 4) = 2\Gamma(n-m)$ for the operator
with root.  It is very essential that the $\Gamma$-functions do not depend on a
and s due to the homogeneity property and depend only on m.  Therefore, the
results obtained are true for any term in the expansion of $\sigma_m$.  Thus,
the DWSG coeffients for the operator with root are expressed through the DWSG
coefficents for the operator without root
\begin{equation}
E_{mr}=\frac{2\Gamma(n-m)}{\Gamma(\frac{n-m}{2})}E_m,
\end{equation}
where
$E_{mr}$ are the DWSG coefficients for the operator with root.  It is
easy to check that $E_0$ and $E_2$ directly calculated in Section 2 for the
operator $\sqrt{-\nabla^2 + V(x)}$ by using the method with the expansion of
root (see formulas (31) and (34)) coincide with the DWSG coefficients given by
the common formula (58).  By using the representation with $e^{\lambda^{p/q}}$,
similarly, it is easy to show that the DWSG coefficents for the operator with an
arbitrary rational root, i.e., for operators of the type $(-\nabla^2 +
V(x))^{p/q}$, where p and q are any natural numbers, are expressed through the
DWSG coefficents for the operator without root as follows:
\begin{equation}
E_{mr}=\frac{q/p\Gamma(q/p\frac{n-m}{2})}{\Gamma(\frac{n-m}{2})}E_m.
\end{equation}
In the particular case of the square root, this formula yields (54) and in the
case of natural root, i.e., $p = 1$, $E_0$ and $E_2$ given by (59) coincide
with $E_0$ and $E_2$ explicitly calculated in Section 3, formulas (43) and (44).
Thus, the problem
of finding of the DWSG coefficients for operators of the
type of rational root of a strictly positive operator is exactly solvable,
i.e., the DWSG coefficients for operators with root are explicitly expressed
through those for operators without root.

Note that it would be of significant interest to calculate the DWSG coefficients
by using two methods proposed for an operator which has eigenmodes.  Finding the
difference between $E_m$ obtained by two methods, we would be able to define
the contribution of eigenmodes to the DWSG coefficients.

\bigskip\bigskip
\section{Acknowledgments}
The author is grateful to V.P. Gusynin and S. Fulling for many valuable remarks
and very fruitful discussions.
The work was supported in part by the grant INTAS-93-2058 "East-West network in
constrained dynamical systems".

\end{document}